\documentclass[conference]{IEEEtran}
\IEEEoverridecommandlockouts
\usepackage{cite}
\usepackage{amsmath,amssymb,amsfonts}
\usepackage{algorithmic}
\usepackage{graphicx}
\usepackage{textcomp}
\usepackage{xcolor}
\usepackage{multirow}
\usepackage[colorlinks,
            linkcolor=blue,
            urlcolor=blue,
            anchorcolor=blue,
            citecolor=blue
            ]{hyperref}
\def\BibTeX{{\rm B\kern-.05em{\sc i\kern-.025em b}\kern-.08em
    T\kern-.1667em\lower.7ex\hbox{E}\kern-.125emX}}
\begin{document}

\title{Prototype: A Keyword Spotting-Based Intelligent Audio SoC for IoT}

\author{Huihong Liang\textsuperscript{1}, Dongxuan Jia\textsuperscript{1}, Youquan Wang\textsuperscript{1}, Longtao Huang\textsuperscript{1}, Shida Zhong\textsuperscript{1,*}, \\
Luping Xiang\textsuperscript{2}, Member, IEEE, Lei Huang\textsuperscript{1}, Senior Member, IEEE, and Tao Yuan\textsuperscript{1}, Member, IEEE \\
\IEEEauthorblockA{\textsuperscript{1}College of Electronics and Information Engineering, Shenzhen University, China \,   \textsuperscript{2}Nanjing University, China \\
\{lianghuihong2022, jiadongxuan2022, wangyouquan2022, huanglongtao2022\}@email.szu.edu.cn, \\
shida.zhong@szu.edu.cn (Corresponding author), luping.xiang@nju.edu.cn, \{lhuang, yuantao\}@szu.edu.cn} 
}

\maketitle

\begin{abstract}
In this demo, we present a compact intelligent audio system-on-chip (SoC) integrated with a keyword spotting accelerator, enabling ultra-low latency, low-power, and low-cost voice interaction in Internet of Things (IoT) devices. Through algorithm-hardware co-design, the system's energy efficiency is maximized. We demonstrate the system’s capabilities through a live FPGA-based prototype, showcasing stable performance and real-time voice interaction for edge intelligence applications.
\end{abstract}


\section{Introduction}
The rapid evolution of IoT and edge computing has driven an exponential increase in demand for intelligent voice systems. In particular, keyword spotting (KWS) plays a vital role in facilitating speech-based interactions with smart devices. Given its real-time and always-on modes, KWS must operate within a stringent power budget while ensuring low latency. Consequently, there is a growing trend towards using FPGA or ASIC to implement KWS to achieve higher energy efficiency. Some KWS accelerators leverage DNNs to push accuracy to higher levels \cite{shan_2023_aad-kws,yang_2024_5-mw}. However, it also leads to increased latency, power and chip area. Integrated smart system-on-chips (SoCs) have also emerged as a key area of research. The combination of CPU, DSP and AI Engine is used in \cite{dong_2023_model-specific} to handle KWS and image classification. However, it suffers from higher latency and chip cost (related to area and power).

To address this challenge, we propose a compact SoC, featuring an energy-efficient KWS accelerator, enabling ultra-low latency, low-power and low-cost voice processing at the edge. Our main contribution lies in the algorithm-hardware co-design, with aggressive algorithm compression and custom hardware computation. In the live demo, we present a fully functional FPGA-based prototype, showcasing the system’s real-time voice interaction capabilities.

\section{System Design}

\begin{figure*}[!t]
\centering
\includegraphics[width=\linewidth]{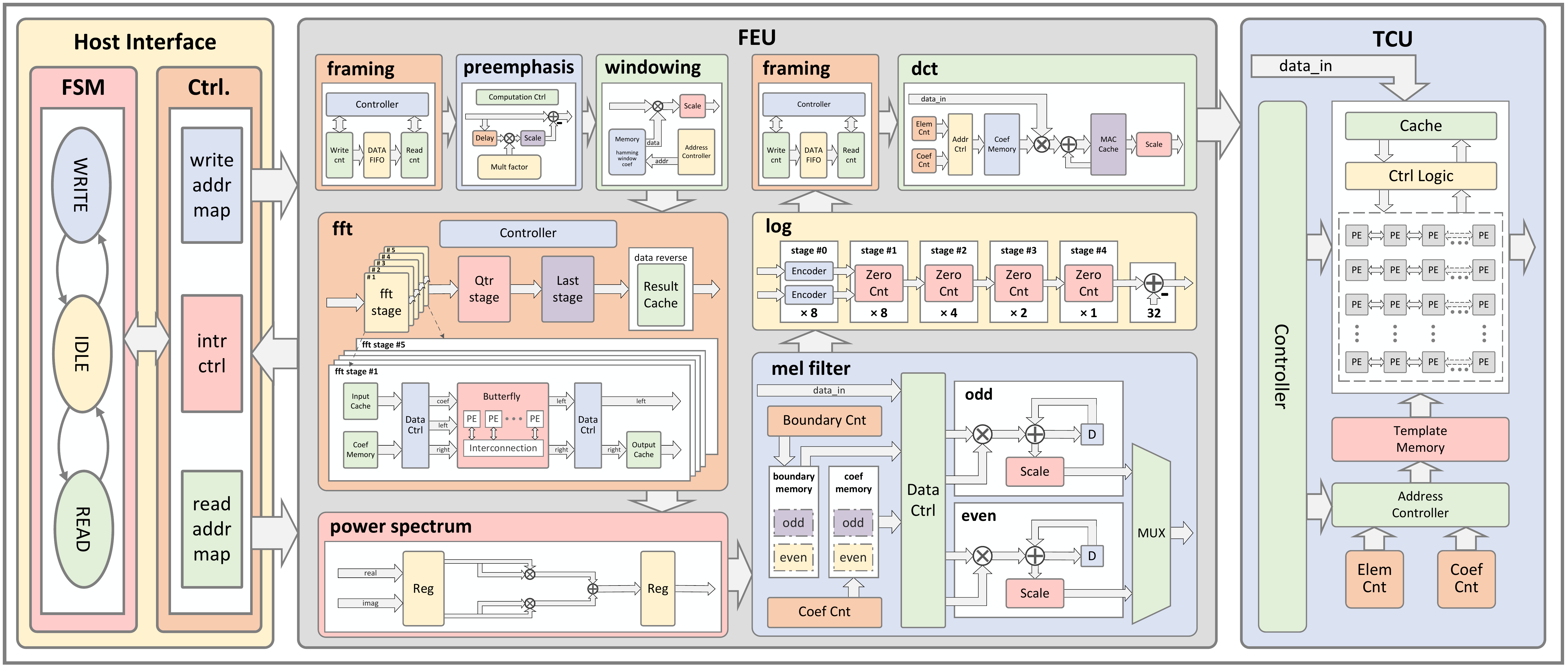}
\caption{Architecture of the proposed KWS accelerator.}
\label{fig_kwsaccelerator}
\vspace{-8.1pt}
\end{figure*}

\begin{figure}[!t]
\centering
\includegraphics[width=\linewidth]{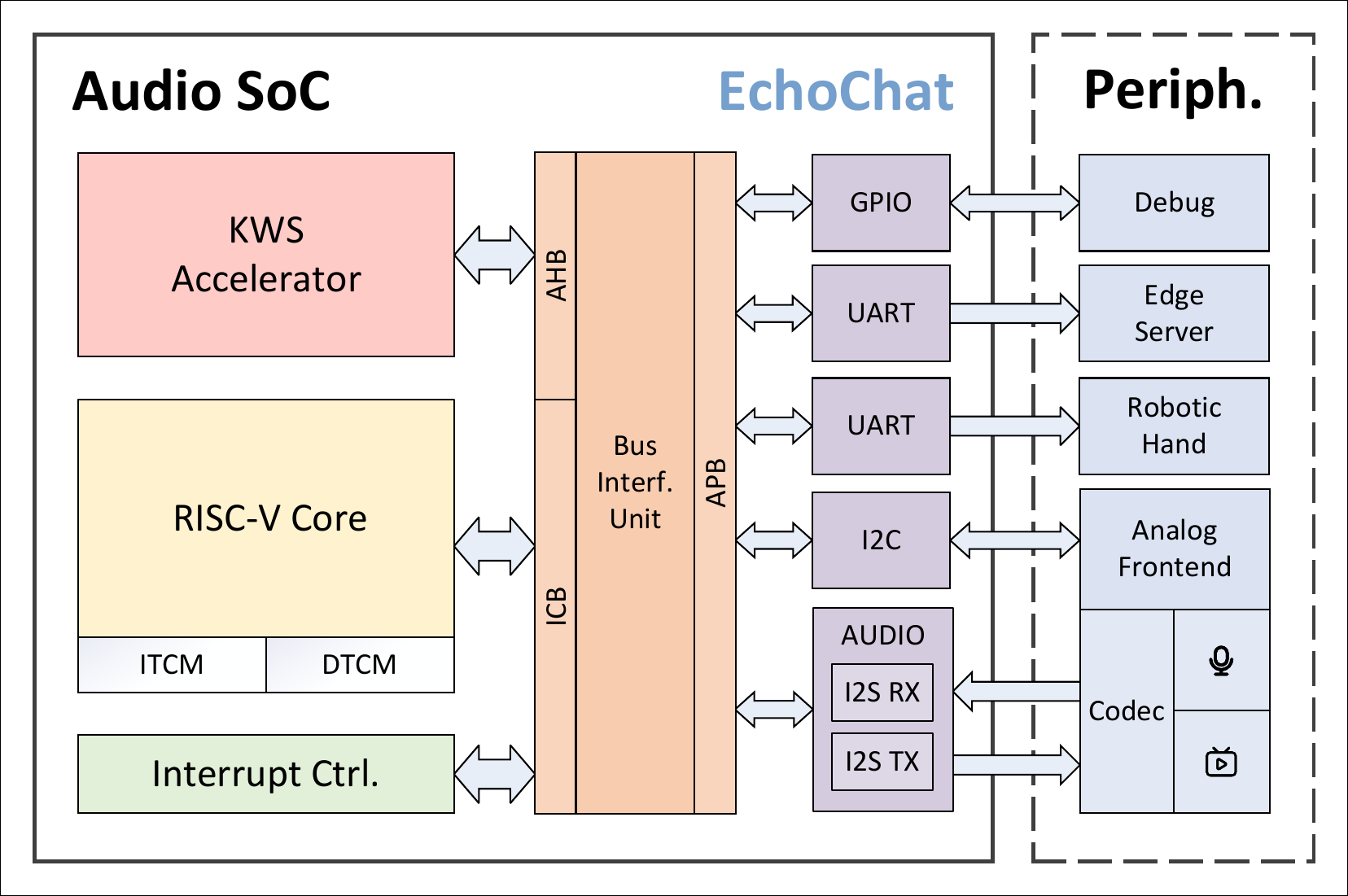}
\caption{Architecture of the proposed audio SoC.}
\label{fig_soc}
\end{figure}

\begin{figure}[!t]
\centering
\includegraphics[width=\linewidth]{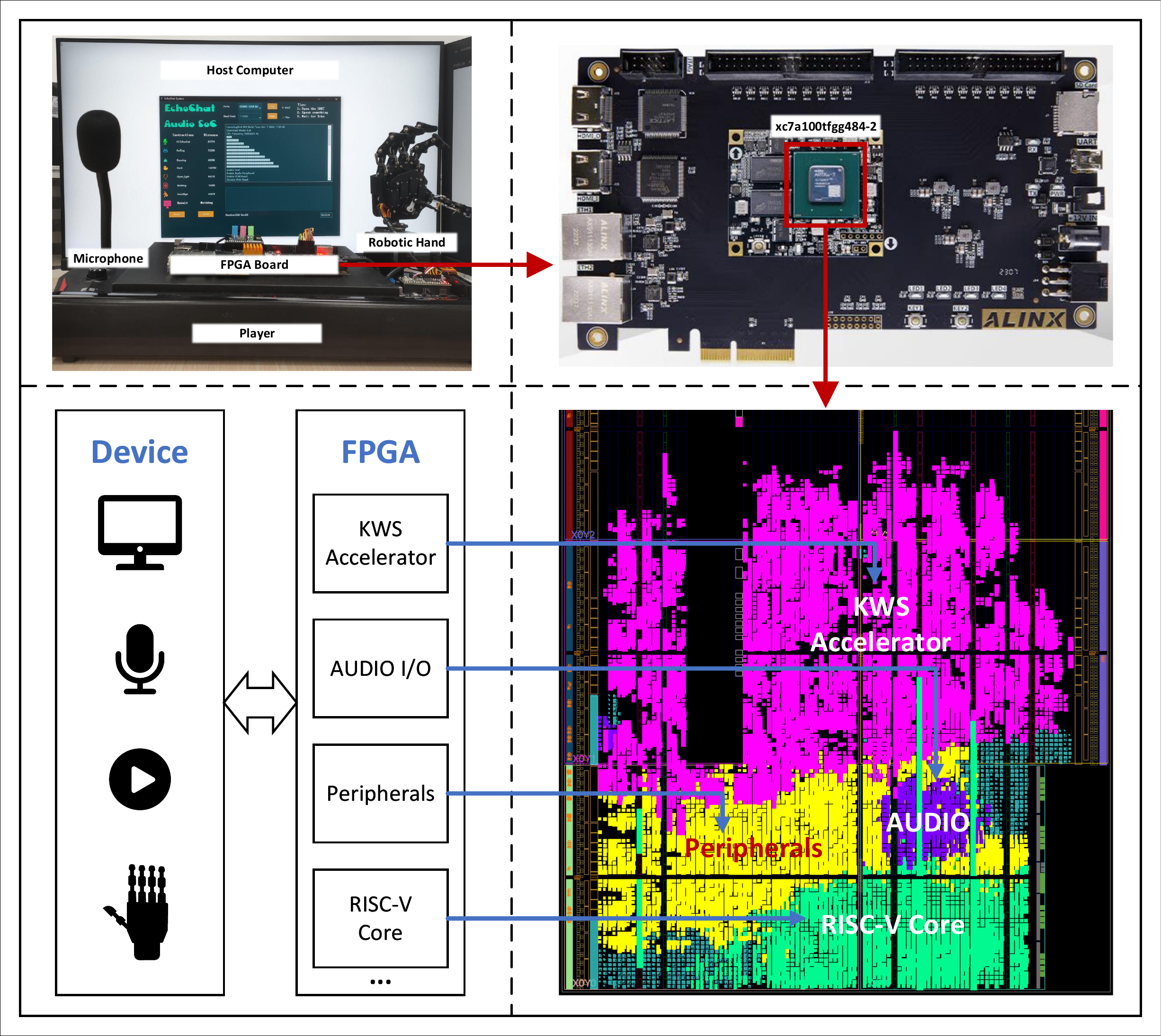}
\caption{Evaluation prototype (top left), FPGA board (top right), illustration of demo (bottom left), and FPGA post-implementation layout (bottom right).}
\label{fig_board}
\end{figure}

\subsection{Algorithm \& Optimization}
Unlike methods bound to large-parameter DNN models, a lightweight approach is adopted in this work, combining Mel frequency cepstral coefficient (MFCC), Vector Quantization (VQ) and Dynamic time warping (DTW). Speech features are extracted using multiple stages of the MFCC pipeline. A codebook is trained using VQ to represent multiple keywords, and template matching is performed using DTW. To compress the algorithm, three strategies are applied: data downsampling, algorithm simplification, and offline parameter optimization, making it suitable for deployment in edge devices.

\subsection{KWS Accelerator}
Based on the optimized algorithm, a dedicated KWS accelerator is designed, as shown in Fig.~\ref{fig_kwsaccelerator}. It consists of two main processing units: the Feature Extraction Unit (FEU) and the Template Classification Unit (TCU), which respectively execute the MFCC and DTW algorithms. Our architectural improvements include: a 7-stage pipeline to address the computational bottleneck of 128-point FFT, sparsity-based Mel filter optimizations, dataflow enhancements for the DCT stages. In the TCU, the dynamic time warping is refined to a fixed diagonal distance computation, reducing area by 99.2\% and power consumption by 84.2\%.

\subsection{Audio SoC}
The KWS accelerator is integrated into the SoC to enable more sophisticated application (Fig.~\ref{fig_soc}). The Nuclei E203 RISC-V is responsible for flexible control, scheduling, and data preprocessing, while the KWS accelerator, serving as a coprocessor, performs speech processing tasks with high parallelism. The custom Audio module is dedicated to driving the Analog Frontend (AFE) for speech input/output. All components collaborate efficiently through an interrupt mechanism.


\section{Evaluation and Demonstration}
Deploying on the E203 RISC-V processor, the optimized algorithm reduces computation time by 98.2\% and memory usage by 98.3\%. Implemented using a 40 nm technology, the SoC operates at 50 MHz and the KWS accelerator at 400 kHz, consuming 12.4 mW and 28.3 $\mu$W, respectively. Compared to KWS works using DNNs \cite{shan_2023_aad-kws,yang_2024_5-mw}, our accelerator achieves a frame latency of 2.98 ms, delivering a 10× speedup over \cite{shan_2023_aad-kws}, with an area of 0.28 mm$^2$, which is 1/6 of the area of \cite{yang_2024_5-mw} after normalization. The SoC has a compact core area of 1.34 mm$^2$ with 128 kB of on-chip memory, further reducing chip cost. Compared to other IoT-oriented audio SoCs \cite{fan_2024_aimmi,dong_2023_model-specific}, our design achieves the smallest area, lowest latency, and just 47.7\% of the power consumption of \cite{dong_2023_model-specific}.

An FPGA-based prototype and evaluation platform are built (Fig.~\ref{fig_board}). The system demonstrates stable performance across over 300 1-second test cases, correctly recognizing up to 89\% of commands, with average processing times of 0.4 ms for KWS and 0.5 s for the SoC. For the live demo, we will showcase the SoC prototype on-site through multiple commands in either English or Chinese, including system wake-up, home appliance control, robotic hand gestures, etc. During demo, the AFE captures and encodes speech, which is processed on the FPGA for command recognition. Results are displayed on the host computer, with responses delivered through the player and executed by the robotic hand and LED. Additionally, components within the prototype collaborate via Bluetooth wireless communication. Through this real-time interactive demo, we hope to provide valuable insights into prototyping edge intelligent systems and inspire further innovation.

\section*{Acknowledgment}
This work was supported by the National Key Research and Development Program, China, under Project No.2023YFB4403805.



\vspace{12pt}

\end{document}